\documentclass{llncs}

\usepackage{hyperref}

\usepackage{graphicx}
\usepackage{amsmath}
\usepackage{mathtools}
\usepackage{url}
\usepackage{multirow}
\usepackage[caption=false]{subfig}
\usepackage{floatrow}
\usepackage{xcolor}
\usepackage{booktabs}
\usepackage{arydshln}
\usepackage{bm}
\usepackage[super]{nth}

\hypersetup{
    unicode=false,          
    pdftoolbar=true,        
    pdfmenubar=true,        
    pdffitwindow=false,     
    pdfstartview={FitH},    
    pdfkeywords={network disruption} {Internet} {latency} {traceroute}, 
    pdfnewwindow=true,      
    colorlinks=true,       
    linkcolor=blue,          
    citecolor=blue,        
    filecolor=magenta,      
    urlcolor=blue           
}

\title{The (thin) Bridges of AS Connectivity: Measuring Dependency using AS Hegemony}

\author{Romain Fontugne\inst{1} \and Anant Shah\inst{2} \and Emile Aben\inst{3} }

\institute{IIJ Research Lab, \email{romain@iij.ad.jp}
\and
Colorado State University, \email{akshah@cs.colostate.edu}
\and
RIPE NCC, \email{emile.aben@ripe.net}}
\begin{document}

\maketitle

\begin{abstract}
Inter-domain routing is a crucial part of the Internet 
designed for arbitrary policies, economical models, and topologies.
This versatility translates into a substantially complex system that is hard to 
comprehend.
Monitoring the inter-domain routing infrastructure is however essential for 
understanding the current state of the Internet and improving it. 
In this paper we design a methodology to answer two simple questions: Which are the 
common transit networks used to reach a certain AS? How much does this AS depends
on these transit networks?
To answer these questions we digest AS paths advertised with the Border Gateway 
Protocol (BGP) into AS graphs and measure node centrality, that is the 
likelihood of an AS to lie on paths between two other ASes.
Our proposal relies solely on the AS hegemony metric, a new way to quantify 
node centrality while taking into account the bias towards the partial view
offered by BGP.
Our analysis using 14 years of BGP data refines our knowledge on Internet flattening
but also exhibits the consolidated position of tier-1 networks in today's IPv4 
and IPv6 Internet.
We also study the connectivity to two content providers (Google and Akamai)
and investigate the AS dependency of networks hosting DNS root servers.
These case studies emphasize the benefits of the proposed method to assist ISPs in 
planning and assessing infrastructure deployment.
\end{abstract}

\section{Introduction}
Networks connected to the Internet are inherently relying on other Autonomous
Systems (ASes) to transmit data.
To determine the path of ASes to go from one place to another,
 the Internet relies solely on the Border Gateway Protocol (BGP).
Computed AS paths are the result of an involved process that considers 
various peering policies set by each connected AS. 
BGP exposes only paths that are favored by ASes hence concealing peering
policies and the exact routing process.
However, as the connectivity of a network depends greatly on the connectivity of the ASes
it relies on, operators need a clear understanding of ASes that are crucial 
to their networks.
Identifying these AS interdependencies facilitates decisions for future deployments,
local routing decisions, and connectivity troubleshooting \cite{wahlisch:pam12}.

In this paper, we aim at estimating the AS interdependencies from BGP data.
We devise a methodology that models ASes interconnections as a graph and 
measure AS centrality, that is the likelihood of an AS to lie on paths between 
two other ASes.
We identify in Section \ref{sec:background} shortcomings of a classical centrality
metric, Betweenness Centrality (BC), when used with BGP data.
From these observations we employ a robust metric to estimate AS centrality, 
called AS hegemony (Section \ref{sec:hegemony}).
We demonstrate the value of the proposed method with 14 years of BGP data (Section 
\ref{sec:results}).
Overall we found that AS interdependencies in IPv4 are decreasing over time which 
corroborate with previous observations of the Internet flattening \cite{comarela:imc16}. 
But we also found that the central role played by tier-1 ISP is reinforced in 
today's Internet.
The Internet flattening for IPv6 is happening at a faster rate, but we found that
Hurricane Electric network is utterly central for the last 9 years.
We also investigated the AS dependency of two popular networks, Akamai and Google,
showing that their dependency to other networks is minimal although their peering policies
are completely different. 
Finally, we look at two networks hosting DNS root servers and show how recent
structural changes to these root servers have affected their AS dependencies.

We make our tools and updated results publicly available \cite{ashash:website}
hence network operators can quickly understand their networks' AS dependencies. 

\section{Background}
\label{sec:background}


\noindent \textbf{Related Work:}
The essence of this work is the estimation of AS centrality in AS graphs.
In the literature AS centrality is commonly measured using Betweenness Centrality (BC).
This is one of the basic metric used to characterize the topology of the 
Internet \cite{zhou:phye04,caida:ccr06}.
It was also applied for similar motivation as ours. 
Karlin et al. \cite{karlin:arxiv09} consider Internet routing at the country-level
to investigate the interdependencies of countries and identify countries relying
on other countries enforcing censorship or wiretapping.
BC is also used to identify critical ASes for industrial and public sectors in 
Germany \cite{wahlisch:pam12}.
Similarly, Schuchard et al. \cite{schuchard:ccs10} select targets for control 
plane attacks using a ranking based on BC.
Finally, researchers have also applied BC to detect changes in the AS-topology.
For example, Liu et al. \cite{rocky:jsac13} employ BC to monitor rerouting events caused by
important disruptive events such as major earthquakes or sea cable faults.
Following these past researches, we initially conducted our experiments using BC 
but faced fundamental shortcomings due to the incomplete view provided by BGP data.
To introduce these challenges let's first review BC.


\noindent \textbf{Betweenness Centrality:}
BC is a fundamental metric that represents the fraction of paths that goes through a node.
Intuitively one expects high BC scores for transit ASes as they occur  
on numerous AS paths, and low BC scores for stub ASes.
Formally, for a graph $G=(V,E)$ composed of a set of nodes $V$ and edges $E$, the 
betweenness centrality is defined as:
\begin{equation}
BC(v) = \frac{1}{S} \sum_{u,w \in V} \sigma_{uw}(v)
\end{equation}
where $\sigma_{uw}(v)$ is the number of paths from $u$ to $w$ passing 
through $v$, and $S$ is the total number of paths.
BC ranges in $[0,1]$, but the relative magnitudes of the scores are usually more 
significant than the absolute values.

\begin{figure}[t]
    \subfloat[Simple graph with three viewpoints (illustrated 
    by looking glasses). The sampled BC and AS hegemony are 
    computed only with best paths from the three viewpoints, the expected BC is 
    computed with all best paths.\label{fig:example}]{\includegraphics[width=.65\columnwidth]{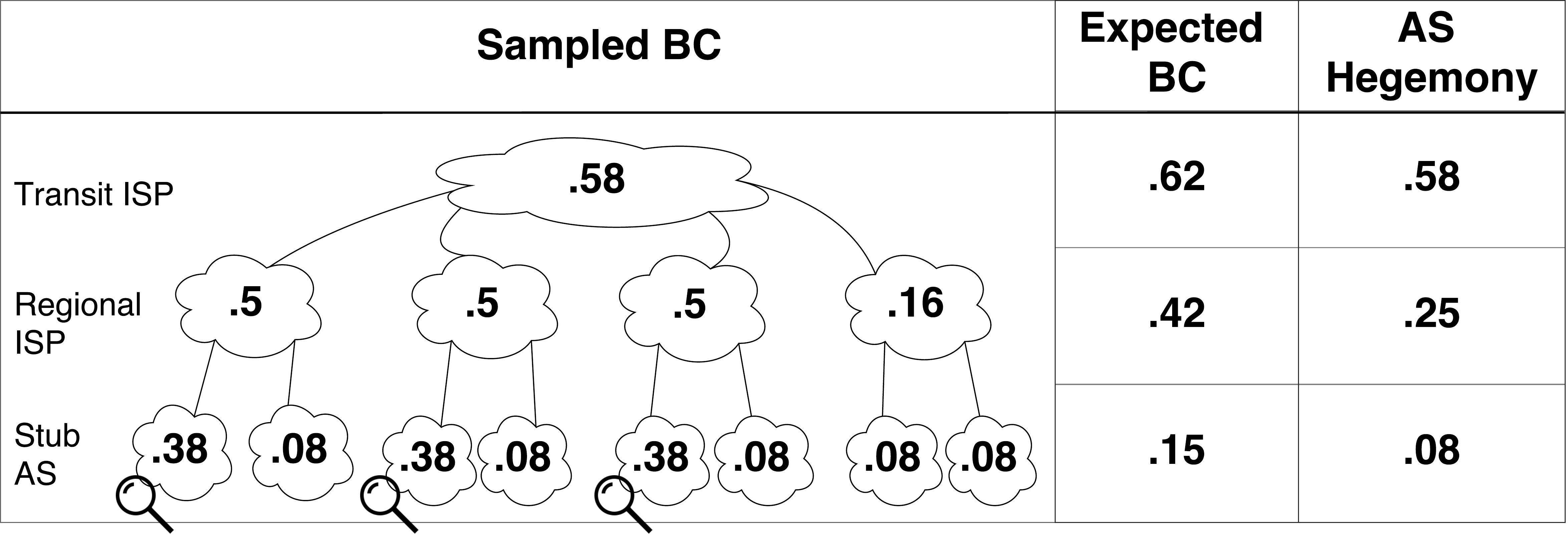}}~~~
    \subfloat[Sampling error forBC and AS hegemony in function of the number of viewpoints.\label{fig:kl}]{\includegraphics[width=.32\textwidth]{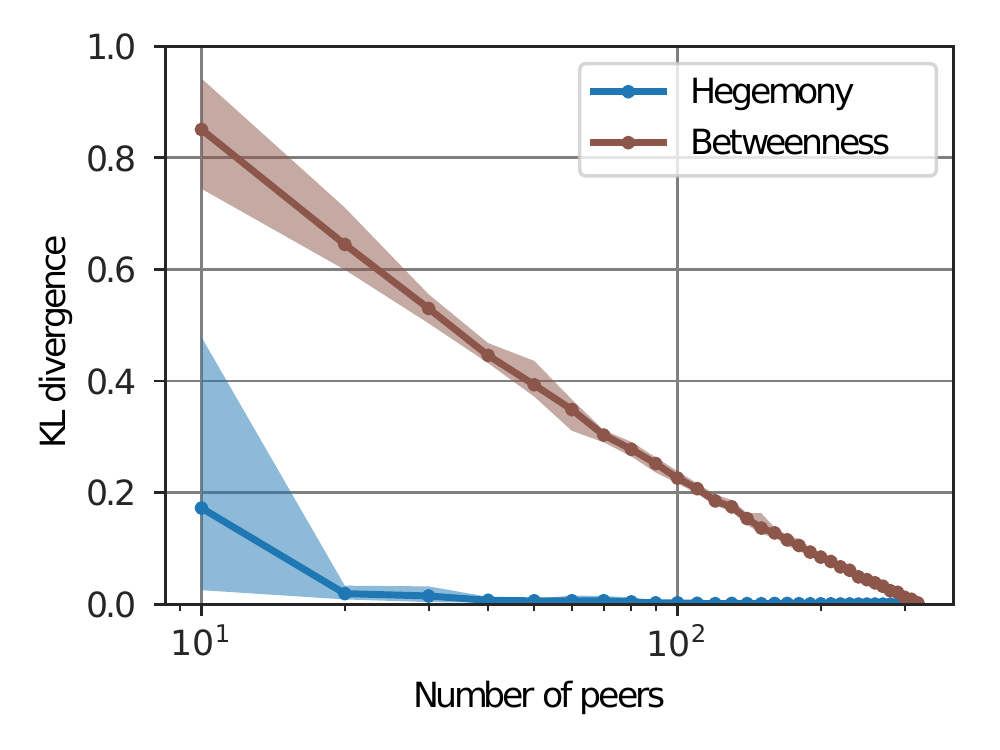}}
    \caption{Comparison of Betweenness Centrality (BC) and AS hegemony with a toy example and BGP data}
\end{figure}


\noindent \textbf{Challenges:}
\label{sec:challenges}
In theory, to compute BC one needs to know the set of all paths in the graph.
With BGP data, however, we are restricted to paths bounded to a small number of viewpoints.
We found that this singular type of path sampling greatly impairs BC results.
To illustrate this, we present an example in Figure \ref{fig:example} with 13 ASes and three viewpoints.
If we had viewpoints in all ASes, thus access to all paths in the graph, we would obtain the highest BC score for 
the transit ISP ($.62$) and lowest scores for the stub ASes ($.15$).
But, using only paths bound to the three viewpoints, the computed BC scores 
are substantially different (Sampled BC in Fig.\ref{fig:example}).
Since about a third of the paths converge to each viewpoint,
BC values for ASes close to the viewpoints are undesirably high 
making these ASes look more central than others. 
This bias is so pronounced that the BC for stub ASes accommodating viewpoints 
($.38$) is twice higher than the BC of one of the regional ISP ($.16$).
Although theoretical studies have already reported that BC is significantly altered by sampling methods
\cite{lee:phyE06}, this issue has been rarely acknowledged in the networking literature.
Mahadevan et al. \cite{caida:ccr06} have reported that BC is not a measure of centrality when computed with network data,
but we stress that this issue comes from the non-random, and opportunistic, sampling 
method used to collect BGP data rather than the metric itself.

In our experiments we construct a global AS graph using all data from the Route Views, RIS, and 
BGPmon project on June \nth{1} 2016. 
This corresponds to an AS graph of 
more than 50k nodes with 326 viewpoints (we consider only full-feed BGP peers), and only $0.6\%$ of all the AS paths on the Internet (16M paths out of the 2.5B).
As collected paths all converge to the 326 viewpoints,
ASes accommodating viewpoints and their neighboring ASes are seemingly more central than other ASes.
To measure the bias obtained with real BGP data we conduct the following experiment.
First, we compute the BC for all ASes from all 326 viewpoints,
then we compare this distribution of BC values 
to BC values obtained with a smaller set of randomly selected viewpoints.
The distance between two distributions is measured with the Kullback-Leibler divergence.
Figure \ref{fig:kl} shows that changing the number of viewpoints invariably 
reshapes the BC distribution, meaning that the obtained BC values are conditioned
by the number of viewpoints.
From these results, we hypothesize that having more than 326 viewpoints would yield
different BC values thus the BC values obtained with the 326 viewpoints might not
be representative of AS centrality.

\section{Methodology}
To address the above BC shortcomings, we devise a monitoring method based
on a robust centrality metric called AS hegemony.
The proposed method consists of two basic steps. 
First we generate graphs from AS paths advertised via BGP. Then, using 
AS hegemony, we estimate the centrality of each AS in the graphs.
We consider two types of graphs, global and local graphs.
They different solely by the scope of the modeled IP space.

\noindent \textbf{Global graph:}
A global graph is made from all AS paths reported by the BGP viewpoints regardless
of the origin AS and announced prefix.
Consequently, these graphs represent the global Internet and central nodes
stand for transit networks that are commonly crossed to reach arbitrary IP addresses.

In 2017, IPv4 global graphs typically contains about 58k nodes and 188k edges 
(about 14k nodes and 43k edges for IPv6). 
The structure of these graphs is complex, yet they are valuable to monitor the 
Internet altogether and reveal routing changes that have Internet-wide impacts.

\noindent \textbf{Local graph:}
A local graph is made only from AS paths with the same origin AS.
Thereby, we compute a local graph for each AS announcing IP space globally. 
Each local graph represents the different ways to reach its origin AS and 
dominant nodes highlight the main transit networks towards only this AS.

These graphs are particularly useful to monitor the dependence of an AS to other
networks. 
In addition, structural changes in local graphs can expose important routing 
changes that are detrimental to the origin AS reachability.

\noindent \textbf{AS Hegemony:}
\label{sec:hegemony}
The core of the proposed method is to quantify the centrality of ASes in the generated graphs.
To circumvent BC sampling problems we extend the recently proposed 
AS hegemony metric~\cite{romain:sigcomm17}.
This metric measures the fraction of paths passing through a node while correcting
for sampling bias.

Computing the hegemony of AS $v$ from AS paths collected from several viewpoints 
consists of the two following steps.
First, AS paths from viewpoints that are bias towards or against AS $v$ are 
discarded.
A viewpoint bias towards AS $v$ means that the viewpoint is located within AS 
$v$, or topologically very close to it, and reports numerous AS paths passing 
through AS $v$.
In contrast, a viewpoint bias against AS $v$ is topologically far from $v$ and is
reporting an usually low number of AS paths containing $v$.
Therefore, viewpoints with an abnormally high, or low, number of paths passing 
through $v$ are discarded and only unbiased viewpoints are selected to compute
the hegemony score.

Second, the centrality of $v$ is computed independently for each unbiased 
viewpoint and these scores are aggregated to give the final AS hegemony value.
That is, for each unbiased viewpoint $j$ the BC of $v$ (hereafter referred as
$BC_{(j)}(v)$) is computed only from AS paths reported by $j$. 
And the average BC value across all unbiased viewpoints is the AS hegemony score 
of $v$.

These steps can formally be summarized into one equation.
Let $n$ be the total number of viewpoints, $[.]$ be the floor function and 
$2\alpha$ be the ratio of disregarded viewpoints. 
The parameter $0\le\alpha<0.5$ represents the ratio of viewpoints that are 
disregarded to compute the hegemony of an AS.
Namely, we discard the top $[\alpha n]$ viewpoints with the highest number of 
paths passing through the AS and do the same for viewpoints with the lowest number of paths.
Then the AS hegemony is defined as:
\begin{equation}
 \mathcal{H}(v,\alpha) = \frac{1}{n-(2[\alpha n])} \sum_{j=[\alpha n]+1}^{n-[\alpha n]} BC_{(j)}(v) 
\end{equation}
where $BC_{(j)}$ is the BC value computed with paths from only one viewpoint $j$ 
(i.e. $BC_{(j)}(v)=1/S \sum_{w \in V} \sigma_{jw}(v)$)
and these values are arranged in ascending order such that $BC_{(1)}(v) \le BC_{(2)}(v) \le \dots \le BC_{(n)}(v)$.

Figure \ref{fig:example} depicts the AS hegemony obtained for the simple graph with three
viewpoints ($\alpha=.34$). 
Unlike the sampled BC, the AS hegemony is consistent for each type 
of node: transit ($\mathcal{H}=0.58$), regional ISP ($\mathcal{H}=0.25$) and stub AS ($\mathcal{H}=0.08$).
AS hegemony scores are intuitively interpreted as the average fraction of paths
crossing a node. 
For example, on average a viewpoint has one fourth of its paths crossing a regional
ISP ($\mathcal{H}=0.25$).

As we did in Section \ref{sec:challenges} with BC, we compute from real BGP data the AS hegemony 
using 326 viewpoints then we compare these results to those obtained with a 
lower number of randomly selected viewpoints.
Figure \ref{fig:kl} shows that the hegemony values with 20 or more viewpoints are very 
similar to the ones obtained from all the peers, hence the AS hegemony is more robust than 
BC to sampling. 
Note that we randomly select peers across different projects (e.g. Route Views, RIS, BGPmon)
to obtain a diverse set of viewpoints. 
Selecting viewpoints from the same BGP collector usually yields poor results \cite{romain:sigcomm17}.

\noindent \textbf{Paths' Weights:}
We also extend AS hegemony to account for path disparities.
In a nutshell, we weight paths according to the amount of IP space they are bound to. 
For example, a path to a /24 IP prefix represents a route to a smaller network than
a path to a /16 IP prefix, thus we give more emphasize to the path towards the /16.
However, the network prefix length alone is not sufficient to resolve the IP
space bound to a path.
Internet address space deaggregation \cite{cittadini:jsac10,julien:cores17} 
should also be taken into account.
For example, a viewpoint reports the path `X Y Z' for the prefix $a.b.c.0/17$ 
and the path `X W Z' for the prefix $a.b.0.0/16$.
Meaning that BGP favors path `X Y Z' for half of the advertised /16.
In this case there is no need to give more emphasis to the path bound to the /16
as each path represents a route to $2^{15}$ IP addresses.

Consequently, we modify our definition of BC to account for the size
of the IP space reachable through a path. 
Formally, $\sigma_{uw}(v)$ is now the number of IP addresses bound to the paths
from $u$ to $w$ and passing through $v$. 
That is the number of IP addresses represented by the advertised IP prefixes minus
the number of IP addresses from covered prefixes (i.e. deaggregated and delegated
prefixes as defined in \cite{cittadini:jsac10}) that are not passing through $v$.
In the rest of the paper this weighted version of BC is applied for the calculation
of AS hegemony in IPv4, but as the relation between number of addresses and 
prefix size in IPv6 is more ambiguous we keep the classical BC definition for
the calculation of AS hegemony in IPv6.


\section{Results}
\label{sec:results}
We have implemented the above methodology in Python and made our tools  and results
publicly available \cite{ashash:website}.
Our implementation uses the BGPStream framework from CAIDA \cite{orsini:imc16} to fetch on the fly 
BGP data and then computes AS hegemony of all ASes in the global graph as
well as AS hegemony for ASes in all local graphs.
We set the parameter $\alpha=0.1$ for all experiments and verified that results 
are consistent with higher $\alpha$ values. 

\noindent \textbf{Dataset:}
The following results are all obtained using BGP data from four BGP collectors,
two from the RouteViews project (route-views2 and LINX) and two from the RIS 
project (rrc00 and rrc10).
These four collectors are selected from the collectors sensitivity results 
presented in \cite{romain:sigcomm17}.
For IPv4 they represent from 51 to 95 BGP peers respectively in 2004 and 2017.
For IPv6, however, as the number of BGP peers is rather small before 2007 (i.e. less than 10 peers)
and AS hegemony values might be irrelevant with such low number of peers (see Fig.~\ref{fig:kl}),
we report only results obtained from 2007 onward using from 11 to 44 peers. 
We obtain Routing Information Base (RIB) files of all peers for the \nth{15} of
each month from January 2004 to September 2017 and compute our methodology on this dataset.

\subsection{IPv4 and IPv6 global graphs}
\begin{figure}[t]
    \subfloat[IPv4\label{fig:longitudinal:4}]{\includegraphics[width=.5\columnwidth]{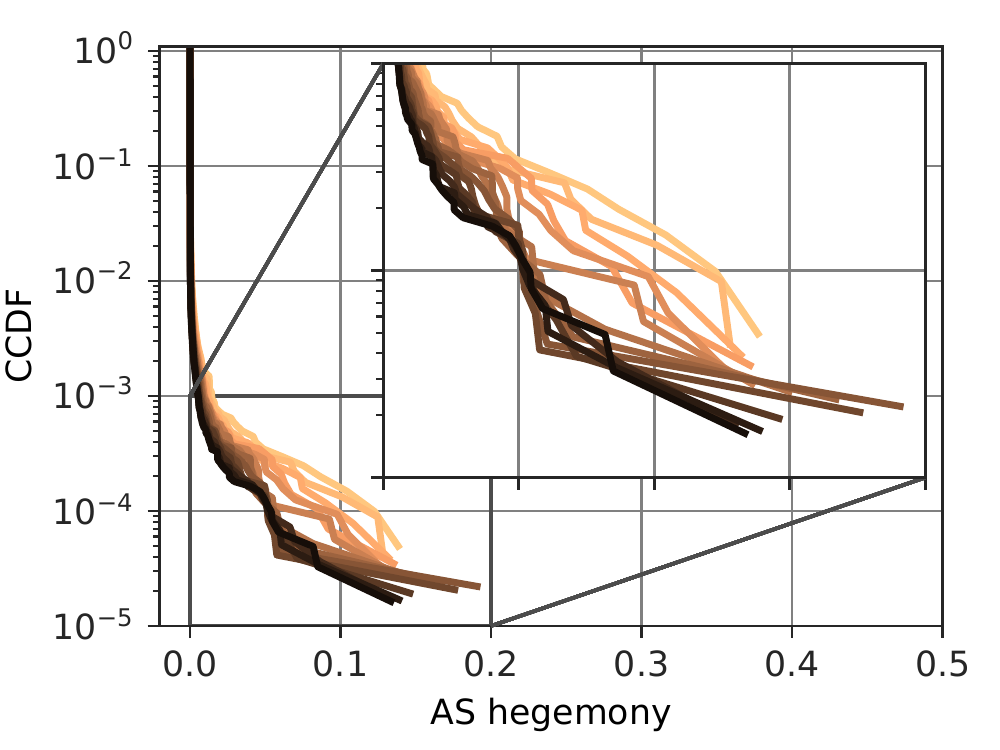} }
    \subfloat[IPv6\label{fig:longitudinal:6}]{\includegraphics[width=.5\columnwidth]{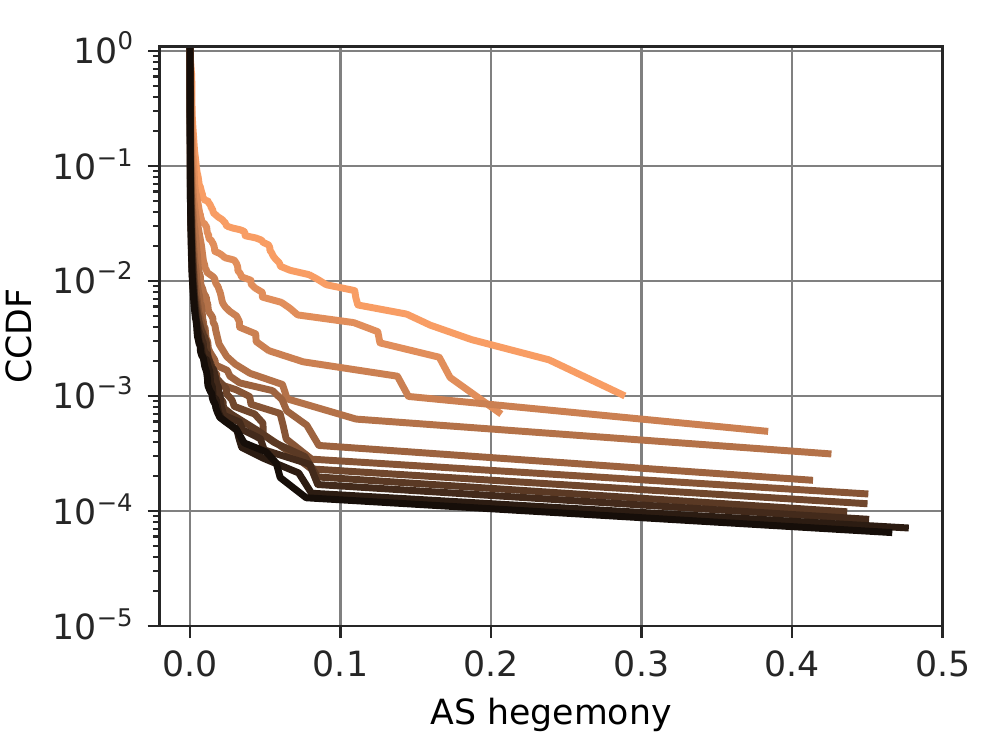}}
    \begin{center}\includegraphics[width=0.5\columnwidth]{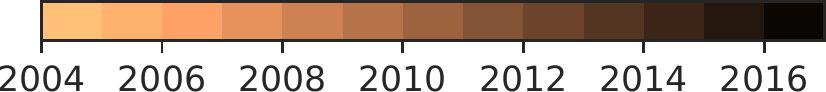}\end{center}
    \caption{Distribution of AS hegemony for all ASes in the global graph.\label{fig:longitudinal}}
\end{figure}

As the starting point of our analysis, we investigate the AS interdependency for the
entire IP space. 
We monitor the evolution of AS hegemony scores in the global AS graph from 2004 to 2017.
Here large AS hegemony scores represent transit networks that are commonly crossed 
to reach arbitrary IP addresses.
Figure \ref{fig:longitudinal} depicts the distribution of the yearly average 
AS hegemony for all ASes in the IPv4  and IPv6 global AS graphs.
In these figures each point represent an AS, and those on the right hand side of
the figures stand for nodes with the highest hegemony values.

As the distribution of AS hegemony values for IPv4
is overall shifting to the left over time (Fig.~\ref{fig:longitudinal:4}), 
we observe a global and steady decrease of AS hegemony values.
This is another evidence of Internet's flattening \cite{comarela:imc16}, 
as networks are peering with more networks we observe less dominant ASes.
Nonetheless, Figure \ref{fig:longitudinal:4} suggests that the AS hegemony for 
the most dominant networks (i.e. points on the right hand side) is quite stable.

\begin{figure}[t]
    \includegraphics[width=\columnwidth]{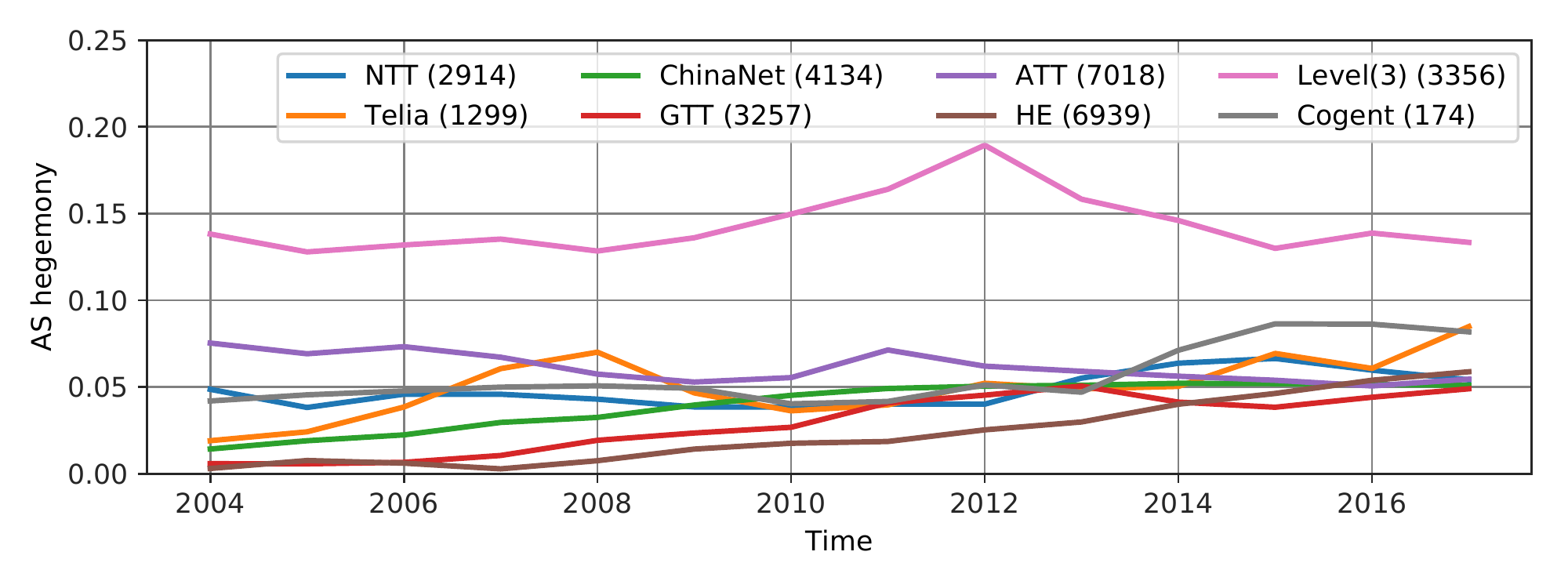}
    \caption{ AS hegemony for Tier-1 ISPs from 2004 to 2017 (global graph, IPv4).\label{fig:tier1}}
\end{figure}
We further investigate this by selecting the eight most dominant ASes found in 
our dataset and monitor their yearly AS hegemony (Fig.~\ref{fig:tier1}). 
The AS hegemony for these networks is indeed either steady, or increasing, which is
contradictory with the global Internet flattening observed earlier.
These two observations provide evidences of dense connectivity at the edge of the
Internet but the role of large transit ISP is still very central to connect
remote places in the Internet.
This can be explained by the growth of public peering facilities (IXP) that allows
local network to keep traffic locally and peer directly with content providers.
Yet transiting to remote locations requires the international networks of tier-1 ISPs.
In recent years this distinction between tier-1 ISP and other networks is event more
visible,
as we observe in Fig. \ref{fig:longitudinal:4} a clear gap between most networks 
($\mathcal{H}<0.03$) and tier-1 ISPs ($\mathcal{H}>0.05$).

Figure \ref{fig:tier1} also depicts the dominance of Level(3) through the entire study period.
After Level(3) acquisition of Global Crossing (AS3549) in 2011, it reached in 2012
the highest AS hegemony score monitored for the IPv4 global graph ($\mathcal{H}=0.19$).
We also found that from 2008 to 2010 Global Crossing was the most dominant
AS in Level(3) local graph, meaning that it was the most common transit network 
to reach Level(3).
These results thus assert that Global Crossing acquisition was the most
effective way for Level(3) to attain new customers.
It also illustrates the benefits of our tools for deployment and business decisions.

For the IPv6 (Fig. \ref{fig:longitudinal:6}) we observe a faster 
Internet flattening than for IPv4. 
We hypothesize that this is mainly because the Internet topology for IPv6 in 
2007 was quite archaic. 
But IPv6 has drastically gained in maturity, the AS hegemony distribution for 
IPv6 in 2017 is then very close to the one for IPv4 in 2009.
The most striking difference with IPv4 is the central role played by
Hurricane Electric in the IPv6 topology.
After doubling its number of peers in 2009 \cite{hui:gigabit09},
Hurricane Electric has been clearly dominating the IPv6 space from 2009 onward.
It reaches an impressive AS hegemony $\mathcal{H}=0.46$ in 2017, largely above the second
and third highest scores (0.07 and 0.05), respectively, for Level(3) and Telia. 
Consequently, our tools confirm the dominant position of Hurricane Electric 
observed previously \cite{dhamdhere:imc12} and permit to systematically quantify 
the overall IPv6 dependency to Hurricane Electric.


\subsection{Case studies: Local graphs}
Our analysis now focuses on results obtained with local graphs.
Unlike the global ones, local graphs shed light to AS dependency only for a specific
origin AS.
We found that the structure of local graphs is very different depending on the
size and peering policies of the origin AS. 
On average in 2017,  an IPv4 local graph contains 98 nodes but 
93\% of these nodes have an hegemony null ($\mathcal{H}=0$).
Typically ASes hosting BGP peers have an hegemony null and AS hegemony
scores increases as the paths converge towards the origin AS.
Thereby, the upstream provider of a single-homed origin AS gets the maximum 
hegemony score, $\mathcal{H}=1$.
By definition the origin AS of each local graph also features $\mathcal{H}=1$, therefore,
we are not reporting the AS hegemony of the origin AS in the following results.

In 2017, local graphs have on average 5 ASes with $\mathcal{H}>0.01$, 
which usually corresponds to a set of upstream providers and tier-1 ASes.
We also noticed interesting graphs containing no dominant AS, and other graphs 
containing numerous nodes with non-negligible AS hegemony scores.
To illustrate this we pick a local graph from both end of the spectrum, namely,
AS20940 from Akamai and AS15169 from Google.

\begin{figure}[t]
    \subfloat[AS20940 Akamai, IPv4\label{fig:akamai:4}]{\includegraphics[width=.5\columnwidth]{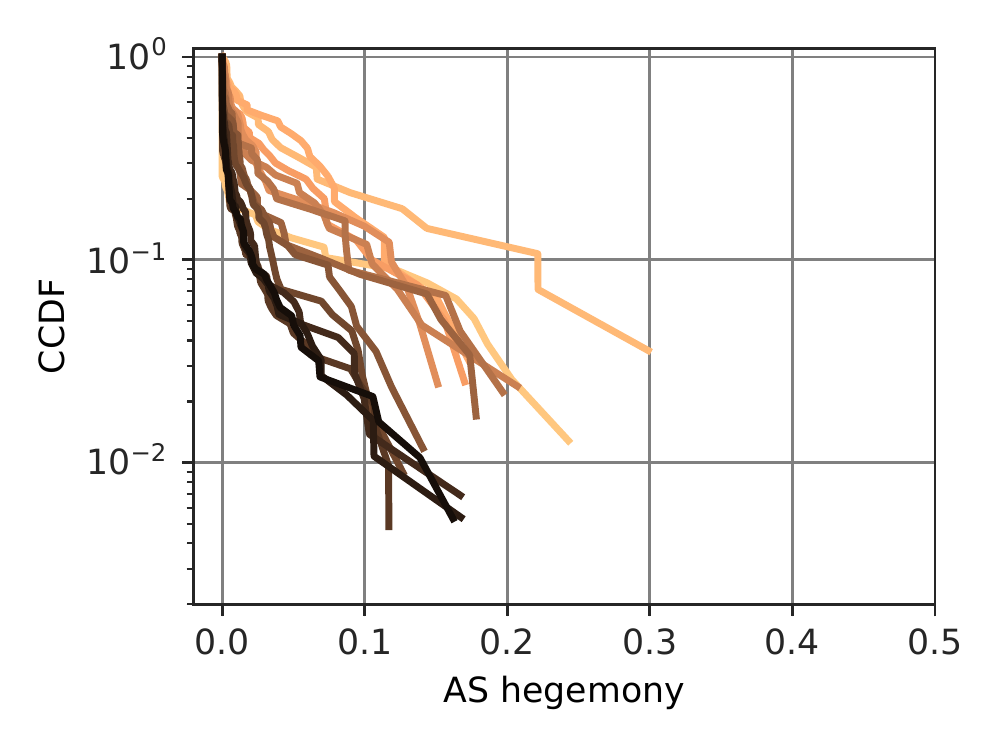}}
    \subfloat[AS15169 Google, IPv4\label{fig:google:4}]{\includegraphics[width=.5\columnwidth]{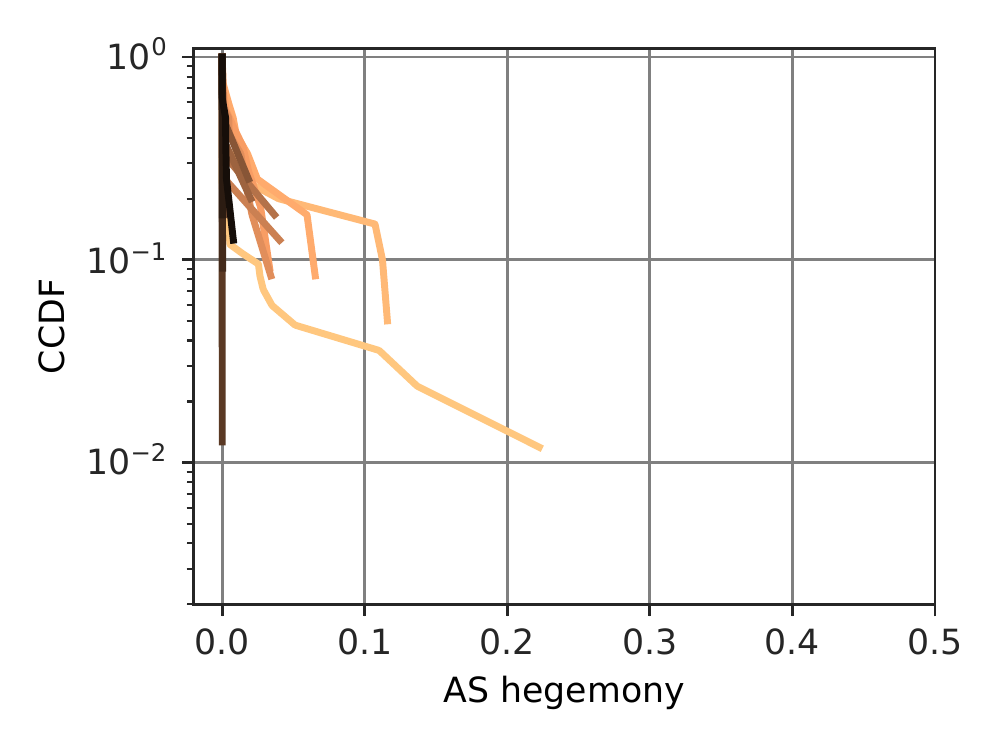}}\\
    \subfloat[AS20940 Akamai, IPv6\label{fig:akamai:6}]{\includegraphics[width=.5\columnwidth]{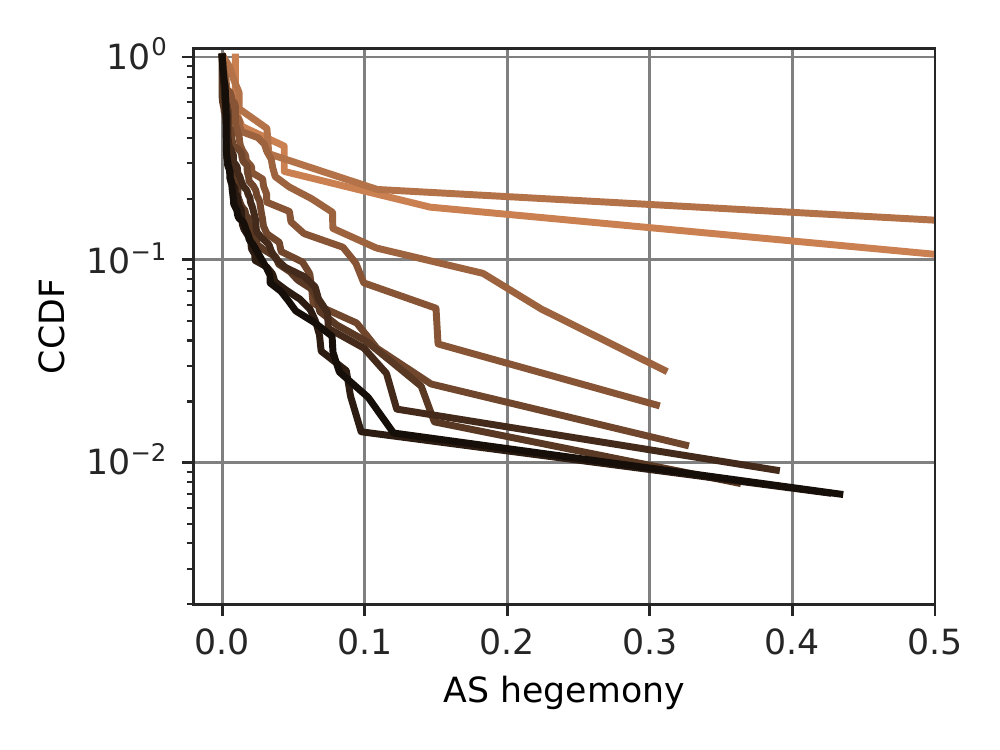}}
    \subfloat[AS15169 Google, IPv6\label{fig:google:6}]{\includegraphics[width=.5\columnwidth]{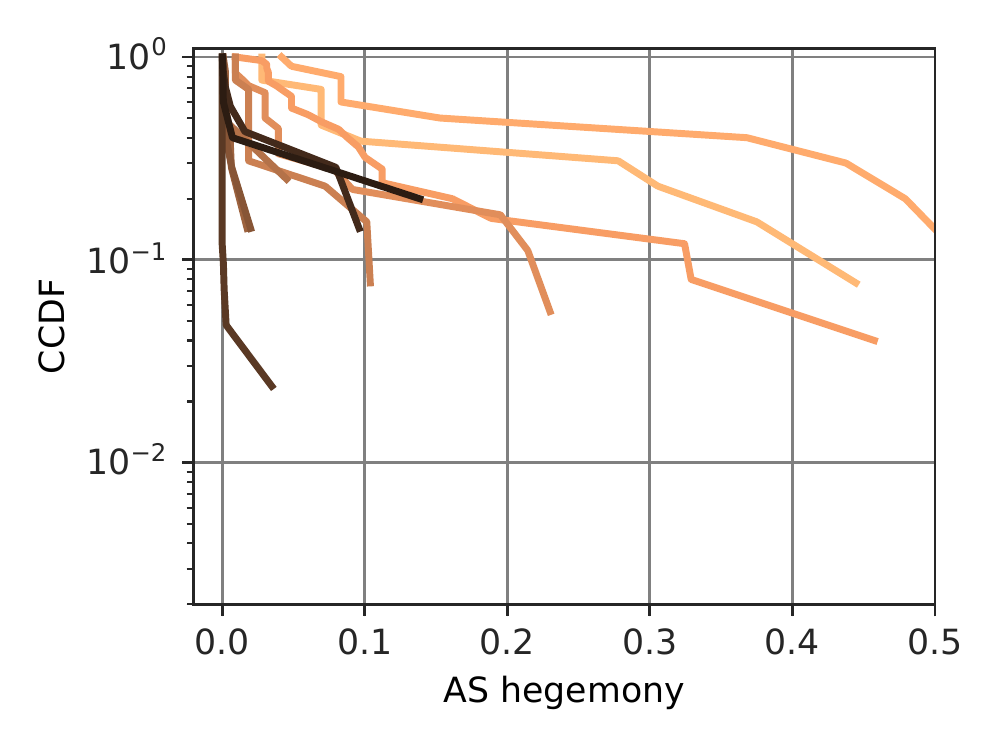}}
    \caption{Distribution of AS hegemony for Google and Akamai local graphs. 
    Same color scale as Fig.~\ref{fig:longitudinal}.}
\end{figure}
\noindent \textbf{Akamai and Google:}
The IPv4 graph for Akamai's main network, AS20940, is the local graph with the 
largest number of nodes in our results.
In 2017, it contains on average 30 nodes with an AS hegemony greater than 0.01
(see Fig.\ref{fig:akamai:4}).
Meaning that accessing Akamai IP space relies on a large set of transit networks.
This is true for our entire analysis period as shown in Figure \ref{fig:akamai:4}. 
Our manual inspection of Akamai BGP announcements reveals that Akamai is 
heavily fragmenting its IP space and advertising small prefixes at various Points 
of Presence (PoPs).
Consequently, each prefix is accessible only through a very limited number of 
upstream providers and all BGP peers report AS paths going through these providers.
In summary, Akamai local graph contains a lot of nodes with weak but non-negligible
AS hegemony scores implying that Akamai has numerous weak AS-dependencies.

On the other hand, the IPv4 graph for Google (AS15169) in 2017 contains no node
with an hegemony greater than 0.01 (see Fig. \ref{fig:google:4}).
Our manual inspection of Google BGP advertisements reveals that, unlike 
Akamai, Google announces all their prefixes at each PoP. 
Because Google is peering at numerous places, all BGP peers report very short
and different AS paths with almost no AS in common hence no relevant hegemony score.
Nonetheless, Google's local graphs before 2012 feature a different AS hegemony distribution
with a few high AS hegemony scores (Fig. \ref{fig:google:4}). 
Level(3) is the most dominant AS observed until 2012.
But then Google has clearly succeeded to bypass Level(3) and alleviate its dependency to
this AS (usually $\mathcal{H}<0.00005$ from 2014), hence now Level(3) is rarely seen in paths towards Google.
In summary, we observe that Google used to depend on a few ASes but it is now 
mostly independent from all ASes.
This is not an isolated case, we have found a few other ASes with no AS dependency,
notably, Microsoft (AS8075), Level(3) (AS3356), Hurricane Electric (AS6939), 
and Verisign (AS7342).

For IPv6, the situations for Akamai and Google is a bit different.
The local graph for Akamai contains a lot of nodes with a high AS hegemony 
(Fig.\ref{fig:akamai:6}). 
But Hurricane Electric is quite outstanding and features an AS hegemony
($\mathcal{H}=0.43$) very close to the one observed
for Hurricane Electric in the IPv6 global graph (Fig. \ref{fig:longitudinal:6}).
Hurricane Electric is also the dominant node in Google's IPv6 local graph
(Fig. \ref{fig:google:6}) but at a much lower magnitude ($\mathcal{H}=0.12$). 
Thereby, our results show that Google's aggressive peering policy has partially 
succeeded to bypass Hurricane Electric IPv6 network.

\noindent \textbf{DNS root servers:}
Monitoring an AS with our tools provides valuable insights into its AS dependency.
This is particularly useful for networks hosting critical infrastructure, as 
operators of these ASes try to minimize their dependencies to third-party networks.
To illustrate the benefits of our tools, we present results for the local graphs 
of ASes hosting DNS root servers. 
Notice that understanding AS dependency of root servers is usually a complicated 
task as most root servers are using anycast and more than 500 instances are deployed 
worldwide.
Due to space constrains, we detail only IPv4 results for networks hosting the 
F-root (AS3557) and B-root (AS394353) servers as they had significant structural 
changes in 2017. 

\begin{figure}[t]
    \subfloat[F-root (AS3557)\label{fig:froot}]{\includegraphics[width=.49\columnwidth]{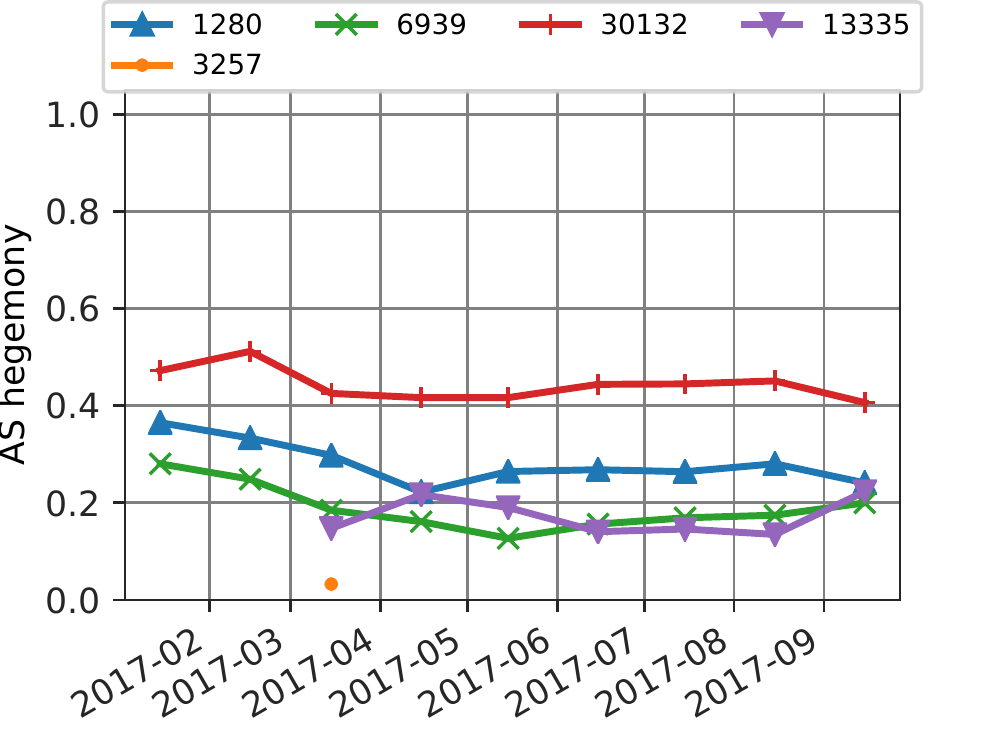}}
    \subfloat[B-root (AS394353)\label{fig:broot}]{\includegraphics[width=.49\columnwidth]{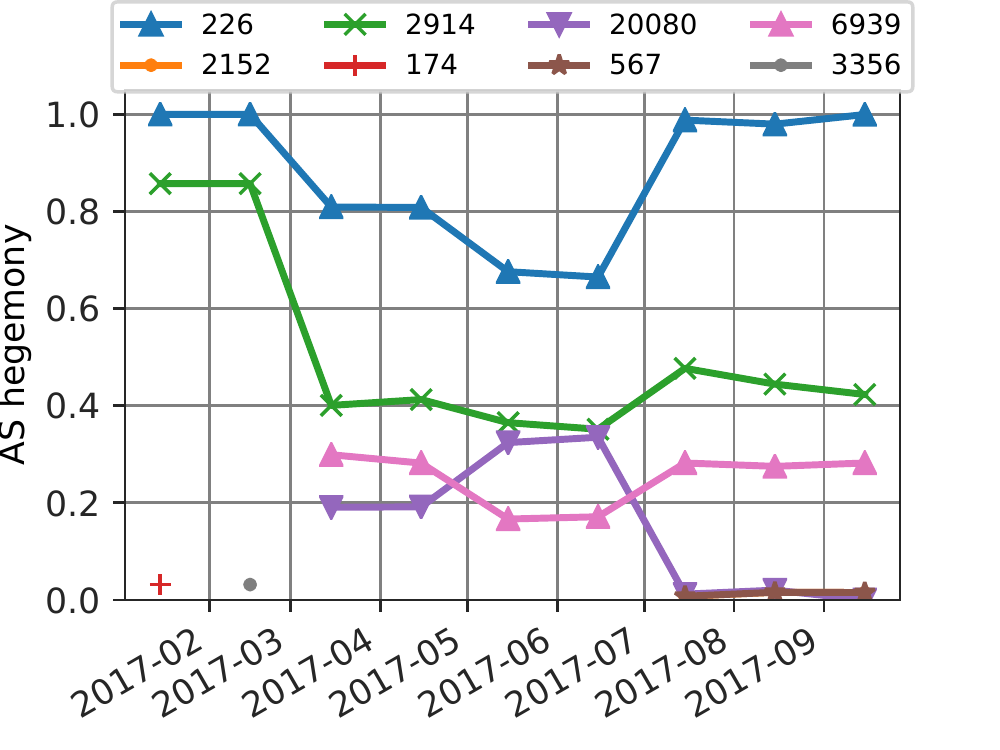}}
    \caption{AS hegemony for nodes in F-root (AS3557) and B-root (AS394353) local graphs from \nth{15} 
    January to \nth{15} September 2017.}
\end{figure}

In early 2017, we observe three dominant transit ASes for the network hosting the F-root server (Fig.~\ref{fig:froot}).
AS30132 and AS1280 are direct upstream networks managed by ISC, the administrator of the F-root server. 
AS6939 is Hurricane Electric, the main provider for AS1280, and is found in about a third of the AS paths toward the F-root server.
From March, Cloudflare (AS13335) starts providing connectivity to new F-root instances~\cite{cloudflare:froot}.
This new infrastructure is clearly visible in our results. 
Starting from March 2017, Cloudflare hegemony is fluctuating around 0.2 and seems to divert traffic from other instances as the three other transit networks have their hegemony proportionally decreased.
From these results we deduce that the addition of Cloudflare has successfully reduced F-root dependencies on other ASes.

For the B-root server (Fig.~\ref{fig:broot}),
we observe two dominant ASes in January and February 2017, Los Nettos (AS226) and NTT America (AS2914). 
Los Nettos reaches $\mathcal{H}=1$ because at that time the B-root server was unicasted and Los Nettos was the sole provider.
NTT also has a very high AS hegemony score, in fact more than 80\% of analyzed AS path also cross NTT's network.
From March 2017, we observe two other transit nodes AMPATH (AS20080) and Hurricane Electric (AS6939).
Our manual inspection of the advertised paths reveals that a single /24 prefix is advertised with AMPATH as the first hop. 
This prefix is one of the two /24 prefixes advertised by the network hosting the B-root server (AS394353) but is not the one containing the server IP address.
We believe that B-root operators were testing anycast in preparation for the 
deployment of the second instance of B-root at Miami that happened in May \cite{broot:anycast}.
In May we acknowledge the deployment of the second instance hosted at AMPATH as the hegemony of that AS is raising again and the one for Los Nettos had significantly decreased.
From July onward, however, we observe a sudden decrease of AMPATH hegemony while hegemony for Los Nettos is getting back close to 1.
Therefore the addition of this second instance had uncertain benefits, 
first, it considerably mitigated the dependence on NTT and Los Nettos networks 
in May and June, but then, from July Los Nettos is once again totally dominating 
the B-root connectivity.


\noindent \textbf{Future directions:}
The structural changes observed for the F and B root servers illustrate the 
value of AS hegemony to monitor significant routing events.
We are now designing an automated detection process to 
identify significant changes in AS hegemony scores. 
This detector reports sudden routing changes such as the recent BGP leak from Google \cite{bgpmon:googleleak}.
During this event Google became a transit provider for NTT OCN, which exhibits 
a sudden and significant increase in Google's AS hegemony for NTT's local graph.
Thanks to AS hegemony detecting this type of event is fairly easy, while state
of the art tools employed by network operators (e.g. BGPmon provided by OpenDNS)
have usually missed this significant event.
As the details and evaluation of this detector go beyond the scope of this paper
we leave this for future work.

In the future we are also planning to investigate different weighting schemes. 
For example by assigning paths' weight based on traffic volume an ISP can
emphasize destinations that are favored by its customers.

\section{Conclusions}
We presented a methodology to quantify the AS interdependency in the Internet.
It deals with the various AS paths reported via BGP and produce AS hegemony scores,
that are robust estimates of the ASes centrality.
Using 14 years of BGP data we proved that this method permits to monitor 
structural changes in the Internet and identify most important ASes to reach a 
certain part of the IP space.
We also demonstrated with case studies the benefits of our tools to help 
ISPs to plan and assess infrastructure deployment.
To assist network operators in these tasks we make our tools and results publicly 
available \cite{ashash:website}.





\bibliographystyle{abbrv}
\bibliography{references}  

\begin{thebibliography}{10}

\bibitem{ashash:website}
{ AS Hegemony Results}.
\newblock \url{http://ihr.iijlab.net/ihr/hegemony/}, 2017.

\bibitem{cittadini:jsac10}
L.~Cittadini, W.~Mühlbauer, S.~Uhlig, R.~Bush, P.~Fran{\c{c}}ois, and
  O.~Maennel.
\newblock Evolution of internet address space deaggregation: Myths and reality.
\newblock {\em IEEE JSAC}, 8(28):1238--1249, 2010.

\bibitem{comarela:imc16}
G.~Comarela, E.~Terzi, and M.~Crovella.
\newblock Detecting unusually-routed ases: Methods and applications.
\newblock In {\em IMC}, pages 445--459. ACM, 2016.

\bibitem{dhamdhere:imc12}
A.~Dhamdhere, M.~Luckie, B.~Huffaker, A.~Elmokashfi, E.~Aben, et~al.
\newblock Measuring the deployment of ipv6: topology, routing and performance.
\newblock In {\em IMC}, pages 537--550. ACM, 2012.

\bibitem{romain:sigcomm17}
R.~Fontugne, A.~Shah, and E.~Aben.
\newblock As hegemony: A robust metric for as centrality.
\newblock In {\em Proceedings of the SIGCOMM Posters and Demos}, pages 48--50.
  ACM, 2017.

\bibitem{julien:cores17}
J.~Gamba, R.~Fontugne, C.~Pelsser, R.~Bush, and E.~Aben.
\newblock Bgp table fragmentation: what \& who?
\newblock In {\em CoRes}, 2017.

\bibitem{cloudflare:froot}
D.~Grant.
\newblock {Delivering Dot }.
\newblock \url{https://blog.cloudflare.com/f-root/}, 2017.

\bibitem{hui:gigabit09}
E.~Hui~Pan.
\newblock {\em Gigabit/ATM Monthly Newsletter November 2009}.
\newblock Information Gatekeepers Inc.

\bibitem{karlin:arxiv09}
J.~Karlin, S.~Forrest, and J.~Rexford.
\newblock Nation-state routing: Censorship, wiretapping, and {BGP}.
\newblock {\em CoRR}, abs/0903.3218, 2009.

\bibitem{lee:phyE06}
S.~H. Lee, P.-J. Kim, and H.~Jeong.
\newblock Statistical properties of sampled networks.
\newblock {\em Phys. Rev. E}, 73:016102, Jan 2006.

\bibitem{rocky:jsac13}
Y.~Liu, X.~Luo, R.~K. Chang, and J.~Su.
\newblock Characterizing inter-domain rerouting by betweenness centrality after
  disruptive events.
\newblock {\em IEEE JSAC}, 31(6):1147--1157, 2013.

\bibitem{caida:ccr06}
P.~Mahadevan, D.~Krioukov, M.~Fomenkov, X.~Dimitropoulos, k.~c. claffy, and
  A.~Vahdat.
\newblock The internet as-level topology: Three data sources and one definitive
  metric.
\newblock {\em SIGCOMM CCR}, 36(1):17--26, Jan. 2006.

\bibitem{orsini:imc16}
C.~Orsini, A.~King, D.~Giordano, V.~Giotsas, and A.~Dainotti.
\newblock Bgpstream: a software framework for live and historical bgp data
  analysis.
\newblock In {\em IMC}, pages 429--444. ACM, 2016.

\bibitem{broot:anycast}
{Root Operators}.
\newblock { B-Root Begins Anycast in May}.
\newblock \url{http://root-servers.org/news/b-root-begins-anycast-in-may.txt},
  2017.

\bibitem{schuchard:ccs10}
M.~Schuchard, A.~Mohaisen, D.~Foo~Kune, N.~Hopper, Y.~Kim, and E.~Y. Vasserman.
\newblock Losing control of the internet: using the data plane to attack the
  control plane.
\newblock In {\em CCS}, pages 726--728. ACM, 2010.

\bibitem{bgpmon:googleleak}
A.~Toonk.
\newblock {BGP leak causing Internet outages in Japan and beyond.}
\newblock
  \url{https://bgpmon.net/bgp-leak-causing-internet-outages-in-japan-and-beyond/},
  August 2017.

\bibitem{wahlisch:pam12}
M.~W{\"a}hlisch, T.~C. Schmidt, M.~de~Br{\"u}n, and T.~H{\"a}berlen.
\newblock Exposing a nation-centric view on the german internet--a change in
  perspective on as-level.
\newblock In {\em PAM}, pages 200--210. Springer, 2012.

\bibitem{zhou:phye04}
S.~Zhou and R.~J. Mondrag{\'o}n.
\newblock Accurately modeling the internet topology.
\newblock {\em Physical Review E}, 70(6):066108, 2004.

\end{thebibliography}

\end{document}